\documentclass[acmsmall]{acmart}

\AtBeginDocument{%
  \providecommand\BibTeX{{%
    \normalfont B\kern-0.5em{\scshape i\kern-0.25em b}\kern-0.8em\TeX}}}

\usepackage{makecell}

\setcopyright{acmlicensed}

\acmJournal{PACMHCI}
\acmVolume{6}
 \acmNumber{GROUP}
\acmArticle{42} 
\acmYear{2022} 
\acmMonth{1}

\acmPrice{15.00}
\acmDOI{10.1145/3492861}

\begin{document}

\title[Designing for Engaging with News using Moral Framing towards Bridging Ideological Divides]{Designing for Engaging with News using Moral Framing towards Bridging Ideological Divides}

\author{Jessica Z. Wang}
\affiliation{%
  \institution{MIT CSAIL}
  \city{Cambridge, MA}
  \country{USA}}
\email{jzwang@mit.edu}

\author{Amy X. Zhang}
\affiliation{%
  \institution{University of Washington}
  \city{Seattle, WA}
  \country{USA}}
\email{axz@cs.uw.edu}

\author{David Karger}
\affiliation{%
  \institution{MIT CSAIL}
  \city{Cambridge, MA}
  \country{USA}}
\email{karger@mit.edu}

\begin{abstract}
Society is showing signs of strong ideological polarization. When pushed to seek perspectives different from their own, people often reject diverse ideas or find them unfathomable. Work has shown that framing controversial issues using the values of the audience can improve understanding of opposing views. In this paper, we present our work designing systems for addressing ideological division through educating U.S.~news consumers to engage using a framework of fundamental human values known as Moral Foundations. We design and implement a series of new features that encourage users to challenge their understanding of opposing views, including annotation of moral frames in news articles, discussion of those frames via inline comments, and recommendations based on relevant moral frames. We describe two versions of features---the first covering a suite of ways to interact with moral framing in news, and the second tailored towards collaborative annotation and discussion. We conduct a field evaluation of each design iteration with 71 participants in total over a period of 6-8 days, finding evidence suggesting users learned to re-frame their discourse in moral values of the opposing side. Our work provides several design considerations for building systems to engage with moral framing.
\end{abstract}

%
%

\ccsdesc[300]{Human-centered computing~Collaborative and social computing}

\keywords{moral framing; social annotation; discussion; online news}

\maketitle

\renewcommand{\shortauthors}{Jessica Z. Wang, Amy X. Zhang, \& David Karger}

\definecolor{added}{rgb}{0, 0, 0}
\definecolor{remove}{rgb}{1, 0, 0}

\section{Introduction}
\textcolor{added}{In today's media landscape, current issues are hotly debated everywhere: on the 24-hour news cycle, in politics, and on social media. Polarization in politics and current issues has been a major area of interest in recent years, with research showing relations between political sides have been getting increasingly acrimonious---Democrats and Republicans' dislike for the opposing party has been increasing~\cite{pew-partisan, iyengar}. While the extent, direction, and nature of trends in polarization is contentious~\cite{poq, iyengar, westfall}, issue-based ideological division (referred to henceforth as ideological division), defined as differences in systems of beliefs on particular issues measured on a traditional left-right scale, remains a source of friction in the political landscape.}

Prior research toward bridging the ideological divide has focused on making diverse content more readily available and on encouraging people to seek diversity~\cite{park2009newscube, 86040d8e37ae4451a215db7cebb07784}. In the past, increased polarization has often been attributed to ideological silos where the same, one-sided sentiments are resonated among a community, so steps have been taken to reintroduce diversity~\cite{pariser2011filter}. However, simply introducing more diverse content or nudging people to seek alternative perspectives is insufficient---studies show that when people are exposed to diverse viewpoints that disagree with their existing beliefs, people may actually double-down on those existing beliefs or dispute the credibility of the source as a mechanism for minimizing their own cognitive dissonance~\cite{nyhan2010corrections}. 
Often, articles of the opposing perspective are not written for audiences of a different ideology from the author. A source of rejection may be that people are resistant to arguments written in the wrong frames, leading to \textcolor{added}{people with} opposing \textcolor{added}{viewpoints} ``talking past each other'' when one party is focused on concerns that the other party is not at all concerned with. \textcolor{added}{Research has shown that it is often the rhetoric, not the facts of the matter, that gives rise to polarized reactions~\cite{lakoff2010moral}.
Therefore, to understand ideological divides in consumption of online news, it is important to examine not only a piece of content's ideological position, but also its \textit{framing}.}
\begin{figure}
\begin{center}
\includegraphics[width=0.50\columnwidth]{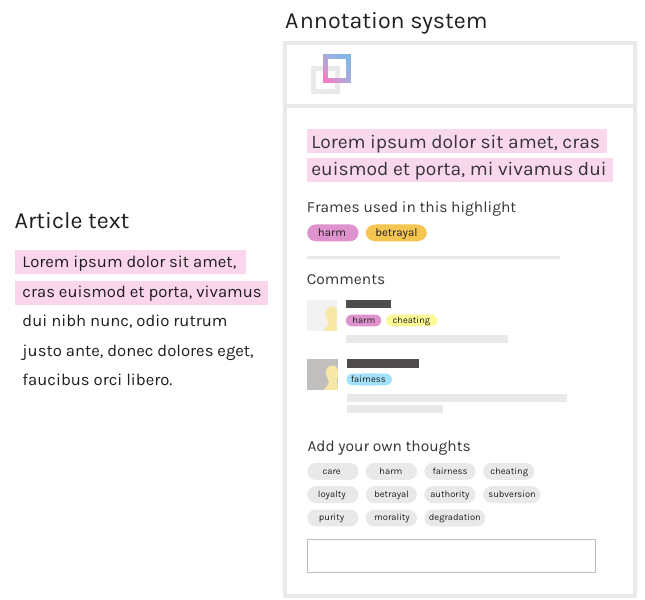}
\caption{System users highlight and annotate article text with moral framing tags, and participate in discussions on the annotations.}
\label{annotation}
\end{center}
\end{figure}
Studies on \emph{moral framing} in particular---framing with a set of fundamental human values---have identified a clear split between the moral values of liberals and conservatives in the U.S.~\cite{graham2012moral, mft2, doi:10.1177/0956797612449177} Many of these studies utilize a moral value framework called \emph{Moral Foundations Theory (MFT)}, which identifies five distinct moral values---fairness, care, sanctity, authority, and loyalty~\cite{mft}. Liberals are distinctly more concerned about values such as fairness and care than conservatives, whereas conservatives are distinctly more concerned about values such as sanctity, authority, and loyalty than liberals~\cite{graham2012moral, mft2}. Past research has demonstrated the power of moral value framing as an especially powerful framework for expressing perspectives to appeal to the opposing side---when liberal arguments on an issue were re-framed with conservative values, conservatives became much more receptive to the liberal viewpoint on the issue and vice versa~\cite{feinberg, doi:10.1177/0956797612449177}.
\textcolor{added}{To leverage the potential of framing and Moral Foundations Theory for news consumption, we introduce our work  designing features for a system that encourages users to use a moral-value based framework when engaging with different perspectives in online content. Our work presents MFT-based treatments on top of familiar social annotation concepts like collaborative annotation and discussion to encourage users to engage with moral framing when engaging with such features and with other users.}

We iterated through two phases of design. For each phase, we designed and developed various features for a system, and conducted a field study evaluation. The first iteration consisted of a large suite of ways to engage with moral framing in news; the second iteration focused in on annotation and discussion. \textcolor{added}{For our field study evaluations, we recruited users interested in reading online news from higher education-affiliated contexts.} Participants were asked to use the system features while reading sets of curated articles for about a week. Our results suggest that compared to a control with users who did not use our system, system users demonstrated positive change in empathy with opposing viewpoints and ability to re-frame arguments into the moral frames of the other side, utilizing more moral value frames in their arguments. Users also indicated that using the system features affected their thinking as they read and discussed issues---engaging with moral framing through the features encouraged them to think about different moral perspectives, as well as the moral values of the people behind those perspectives.\\
In summary, the contributions of this paper are: (i) the design of a series of features for engagement with news using moral framing; (ii) two field study evaluations demonstrating the effectiveness of our approach on utilizing framing in political discussions, as well as empathy toward counter-attitudinal perspectives; and (iii) takeaways for future moral framing-related work in the online media and discourse space.

\section{Background and Related Work}
We first discuss the importance of framing towards crafting messages to bridge ideological divisions; then, we describe how our work contributes to the class of news systems for promoting ideological diversity. Finally, we situate our work at the intersection of tools to illuminate media frames and social annotation systems for education.

\subsection{Echo Chambers and Selective Exposure}
Political theorists agree that exposure to information that challenges one's preconceived views and opinions is necessary to develop accurate beliefs~\cite{frey, pmid19586162}. Diverse content consumption, i.e., a balanced news diet, creates more informed societies, healthier democracies, and a more solid understanding of one's own beliefs~\cite{marwick2017media}. 

Unfortunately, research has shown that people consuming content or engaging with people and things around them suffer from \emph{selective exposure}, a preference for seeking out information that supports our existing beliefs and avoiding contradictory perspectives. Selective exposure reinforces the existence of echo chambers. When people selectively expose themselves to content online and exist within echo chambers, they do not get a diversity of perspectives~\cite{selectiveexposure}. 

Our unconscious tendency to selectively expose in the first place is explained by \emph{cognitive dissonance}, the state of mental stress and discomfort experienced by those holding inconsistent or conflicting thoughts~\cite{festinger1962theory}. When given the power of choice, people tend to give preference to information that conforms to existing mental models, over new, conflicting information~\cite{kastenmuller2010selective}.

The problem with the failure to engage with diverse perspectives in news consumption is twofold: first, it is difficult to get diverse news in front of people; second, once people finally do encounter diverse news content, it is difficult to motivate them to engage productively.

\subsection{Systems to Promote Availability of Diverse Content}
A number of systems have been built to tackle the first problem of encouraging or even directly introducing greater diversity into online social spaces. Recommendation algorithms modify the content one sees---Sidelines increases diversity in news and opinion aggregators by not allowing the preferences of any one user group to dominate the results~\cite{munson}, and Garimella et al. proposed an algorithm to determine who to best target with opposing-view content~\cite{DBLP:journals/corr/GarimellaMGM16}. 
News feeds such as NewsCube~\cite{park2009newscube} and Blue Feed, Red Feed~\cite{bluefeed}, sites such as AllSides~\cite{allsides}, as well as discussion systems like Opinion Space~\cite{faridani2010opinion} and ConsiderIt~\cite{Kriplean:2012:SRP:2145204.2145249} show diverse content in the context of each other, inviting readers to explore multiple and contrasting viewpoints.

However, research has shown that while some users may be diversity-seeking, others remain challenge-averse in the face of diverse content~\cite{munson2010presenting}. As a result, systems have been designed to nudge people towards more diverse reading. 
For instance, behavioral insight systems such as Balancer~\cite{86040d8e37ae4451a215db7cebb07784} and FollowBias~\cite{matias2017followbias} bring visibility to the imbalance of perspectives within a user's habits or social network.

\subsection{Systems to Motivate Consumption of Diverse Content}
Once people finally do encounter diverse content, it can be difficult to engage productively. A person reading diverse content that presents views opposite to their own is likely not the intended audience of the writing. Research shows that exposure to opposing views on social media can in some occasions increase political polarization   \cite{Bail9216,nyhan2010corrections}. Kim et al.~found that in difficult online discussion, users are extremely wary of system interventions that may backfire. However, they also found that in online conflict, users were excited by technology that helps them reach more common ground with others on the opposite side \cite{kimetal}. 

In utilizing user commonalities or different framing for motivating consumption of diverse content, some work aims to alter search engines towards recommending articles that have a language model closer to the reader's own use of language~\cite{yom2014promoting}. Other work aims to alter the presentation of opposing content to encourage mindful engagement with opposing perspectives. For instance, social cue systems aim to surface commonalities between otherwise contrasting types of users. Some work suggests that people with opposing views might come together over other shared common interests~\cite{graells2014people}. However, when Agapie \& Munson added social annotations such as location, friendships, employer, etc. to show users that a counter-attitudinal story was shared by people similar to them~\cite{agapie2015social}, these signals mostly failed to alter reader interest.
Another study found that stranger annotations had no impact while friend annotations increased reader engagement and interest~\cite{kulkarni2013all}. However, friend annotations may be less effective towards encouraging diverse news reading because of homophily, the tendency for people to associate with people who are like them.
Finally, Liao et al. studied using source position indicators on others' opinions in online discussions~\cite{Liao:2014:YHM:2531602.2531711}, suggesting that people's degree of interest in diverse content depends on situational factors and the domain or topic.

Still, other work invites readers to deeper reflection and consideration of opposing perspectives while engaging with content.
For instance, users can go beyond exploration in Opinion Space~\cite{faridani2010opinion}
and ConsiderIt~\cite{Kriplean:2012:SRP:2145204.2145249} towards writing and annotating content blending multiple perspectives.
Similarly, systems like Reflect~\cite{kriplean2012you} and Wikum~\cite{zhang2017wikum} ask readers to summarize another's or multiple viewpoints to arrive at a more holistic understanding of different perspectives.
Research on Wikipedia editors have also found that their edits become less slanted over time~\cite{greenstein2016ideological}.  Even more recent work further demonstrated that in contributions to Wikipedia articles, contributors' polarization of views declined over time \emph{because} contributors were writing content with opposing perspectives, and self-moderating their own contributions over time \cite{greenstein}. Additionally, Warner et al. demonstrated that first-person perspective-taking and common ingroup cooperative narrative-writing exercises improved affect and perceived similarity with people on the other political side \cite{warner}.  

Our work builds directly on prior work on both passive annotations and active engagement with content. Over the course of two iterations of our system, we incorporate several features to encourage exploration and understanding of diverse perspectives, including recommendations, annotations, commenting, and summarization. Unlike prior work, we focus on engagement with underlying moral frames as opposed to differing stances.

\subsection{Moral Foundations Theory}
Our research into moral framing draws from Moral Foundations Theory (MFT)~\cite{mft}, stating that human moral thought is based upon five fundamental psychological foundations that are found across humanity. 
These five foundations are:

\begin{itemize}
    \item Care/harm. This foundation is related to an ability to feel (and dislike) the pain of others. It underlies virtues of kindness, gentleness, and nurturance.
    \item Fairness/cheating. This foundation generates ideas of justice, rights, and autonomy.
    \item Loyalty/betrayal.	This foundation underlies virtues of patriotism and self-sacrifice for the group.
    \item Authority/subversion.	This foundation underlies virtues of leadership and followership, including deference to legitimate authority and respect for traditions.
    \item Sanctity/degradation.	This foundation underlies notions of striving to live in an elevated, less carnal, more noble way.
\end{itemize}

Research shows that liberals and conservatives \textcolor{added}{in the U.S.} have distinctly differing moral foundations. Liberals prioritize care and fairness more than conservatives, whereas conservatives prioritize authority, sanctity, and loyalty more than liberals~\cite{mft2}. This work is also echoed by the work of George Lakoff~\cite{lakoff2010moral}, characterizing conservatives' arguments as using metaphors that invoke a ``strict father model'', while liberal arguments operate under a ``nurturant parent model''.
People from one side can be swayed to support policies from the other side if the policies are framed using the moral foundations they care about. For instance, researchers took a traditionally conservative argument framed in terms of loyalty (``English should be the official language of the U.S. because it unites Americans and is a fundamental part of a larger cultural assimilation process'') then re-framed it using a liberal moral value of fairness (``making English the official language of the United States would compel immigrants to learn the language, leading them in turn to face better job prospects and less discrimination''). The re-framed statement was uniquely persuasive to both \textcolor{added}{U.S.} liberals as well as those who simply highly valued fairness~\cite{feinberg}. Another study was done in the context of environmental science, where re-framing pro-environmental rhetoric in terms of the conservative moral value of purity largely eliminated the difference between \textcolor{added}{U.S.} liberals' and conservatives' environmental attitudes~\cite{doi:10.1177/0956797612449177}. Differences in underlying moral foundations may be key to understanding why arguments framed in the moral values of one side may fall on deaf ears when presented to the opposing side. 

In light of the potential in utilizing moral framing, we consider in this work alternative means by which we can communicate and teach moral framing to users.

\subsection{Systems and Literature towards Recognizing Framing}
\label{prior-work-framing}
A separate line of work exists around information literacy and recognizing media framing. Research on visualizations to explore media framing~\cite{diakopoulos2015compare} and moral framing in blogs~\cite{diakopoulos2014identifying} seek to engage journalists to interrogate media frames. Initiatives such as GlobalVoices NewsFrames~\cite{newsframes} have provided tools such as Media Cloud~\cite{mediacloud} that empower people to interrogate frames, enabling work such as investigating the media ecosystem around content generated on issues~\cite{FM4947}. Social media has also been a source of content for researchers' analysis of media frames and the ways they are used in furthering discourse on current events~\cite{Stewart:2017:DLC:3171581.3134920}.

However, there is little work aimed at teaching \emph{readers} to recognize media frames or encouraging participants in discussion systems to engage with framing in their arguments. Recently, Jonathan Haidt, one of the creators of MFT, and others have developed the Viewpoint Diversity Experience (VDE)~\cite{heterodox}, a set of videos, articles, and other literature aiming to teach about differing moral frames towards promoting dialogue across differing viewpoints. However, a limitation of the VDE is that it is an intensive classroom-style course requiring dedicated time. Our work aims to build upon this work by using design interventions towards the goal of teaching readers about moral foundations and recognizing moral framing in a way that is applicable to peoples' everyday life. 

\subsection{Social Annotations for Improved Engagement}
\label{social-annotation}
As influential as news is, so too is the means by which we interact with it as an educational medium. Enter social annotation (SA)---the ability for users to modify content with highlights, comments, and other content-relevant or contextual information. \textcolor{added}{While much prior work points to the incivility of public SA systems such as open comments sections, suggesting such systems may polarize users~\cite{nasty}, SA as a means of interacting and engaging with content has been shown to be a particularly effective educational tool~\cite{Zyto:2012:SCD:2207676.2208326}, with various SA systems deployed in live classrooms today.}

Social annotation boosts participation and engagement~\cite{lebow, leetiernan}, improves instruction~\cite{lebow}, promotes attention, organization, and communication~\cite{yang}, and improves reading comprehension skills~\cite{archibald}. Furthermore, other readers benefit from social annotations present on a page---these future readers get exposure to new ideas, see other perspectives, and build knowledge about the annotated content~\cite{kawase}. Social annotation is also an application of the principles from Bloom's taxonomy, utilizing the three most effective methods of learning: \emph{analysis}, \emph{evaluation}, and \emph{creation}~\cite{bloom}.

Similarly to social annotation, collective summarization shows promise as an effective intervention for improving user engagement with content. Wikum+, a system which centers social discussion and summarization, helped groups organize and collaborate better \cite{Tian2020ASF}.




\section{Designing System Features for Engaging With Moral Framing}

Building on the two research areas of political psychology literature on moral framing and computer-mediated design research on encouraging diverse reading and engagement with opposing \textcolor{added}{viewpoints}, we present the design of a set of features to help people consume and think about news in a moral framework. Establishing such a framework lays a common ground upon which readers can connect to the perspectives of articles they may not agree with and participate in more constructive, rather than destructive, discussion. To this effect, our main design objective is to prompt users to think about moral framing as they read articles and participate in discussions around articles. \textcolor{added}{Our work explores whether there is potential to encourage deeper engagement with perspectives and framing if we augment existing familiar social engagement concepts with treatments based on Moral Foundations. These features are not necessarily meant to exist in the global public domain, but instead should be considered in the context of online communities that have potential for genuine and productive discourse.} 

Research suggests that people overestimate the ideological extremity of the opposing side's moral concerns~\cite{graham2012moral}. In using our features, users are empowered to uncover the actual usage of moral frames in an argument. Observing moral values that one agrees with can create mutual understanding and connection toward facilitating productive conversation; observing divergent moral values from one's own can bring awareness to why people \textcolor{added}{with opposing viewpoints} may talk past one another.

We iterated through two versions of design. In the first iteration, we created a suite of features for interacting with moral framing in news. These features include 1) in-page annotation: highlighting, annotating, tagging, and commenting using the moral foundations framework; 2) page-level summarization of moral frames present, 3) recommended articles with moral framing tags; and 4) Wiki-like collaborative article summaries. In the second iteration of the system, and our latest version, we focused on in-page annotation, with a greater emphasis on commenting and facilitating discussions. \textcolor{added}{For both iterations, we conducted a user study that brought together users with potentially differing moral foundations on several current issues in the news to evaluate each design and collect qualitative insights.}

\subsection{Implementation}
Our system consists of a Chrome extension client application in Javascript, HTML, and CSS, and a Django back-end server with a Postgres database. On the client side, the app injects content scripts into article pages that enable highlighting and annotation features. This communicates with the server to both store and fetch user highlights, tags, and comments. The extension pop-up interface consists of a set of tabs where the page framing, recommendation, and summary features live. The AYLIEN news API~\cite{aylien} powers the recommended articles feature.

\section{Iteration 1: V1}
In this section, we describe the design and evaluation of the first design iteration.

\subsection{Design}
The first iteration of these features sought to surface moral framing and introduce opportunities for users to engage with said moral framing through a number of features that would raise awareness of framing to users while consuming news articles. Our design aimed to be lightweight and minimally disruptive to a user's browsing experience, yet intriguing enough to encourage users to engage with moral framing. We describe the two forms of user interaction with the features in the following sections.

\subsubsection{In-page highlighting and annotation}
\label{inpagehighlight}
\begin{figure}
\begin{center}
\frame{\includegraphics[width=0.40\columnwidth]{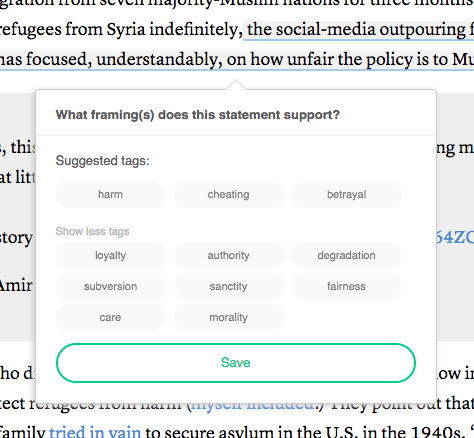}}
\hspace{0.2cm}
\frame{\includegraphics[width=0.40\columnwidth]{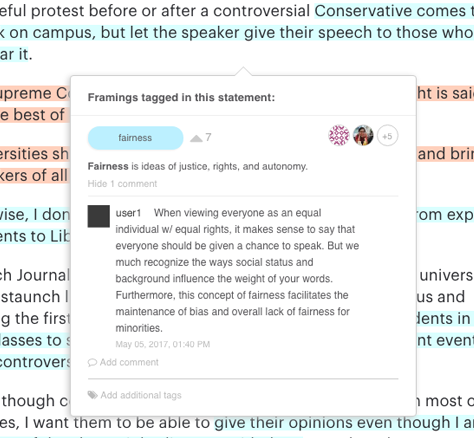}}
\caption{Features from V1 of the system. Left: frame-tagging interface; right: annotation interface.}
\label{panov1}
\end{center}
\end{figure}

The first user interface \textcolor{added}{feature} provided the ability to highlight and annotate news articles across the web as seen in Figure \ref{panov1}. This feature lived inline on article pages themselves. Users reading a news article could select via highlighting a sentence or passage, enabling a pop-up bubble to appear inline. Clicking on the pop-up triggered an interface to appear directly on the page below the selected text that prompts the user: ``What framing(s) does this statement support?'' Users then selected the appropriate moral framing tags and saved the highlight, creating an inline annotation that became visible to all users.

This feature generated suggested moral frame tags for each article as a starting point for users. In order to automatically detect which moral values are likely to be present, we used the Linguistic Inquiry and Word Count (LIWC)~\cite{liwc} approach with a supplementary Moral Foundations dictionary developed by Graham et. al.~\cite{mftdic}. In addition to the 10 moral foundations, the LIWC dictionary included an additional moral frame of \emph{morality}, or appeal to a general human moral system.

\emph{The highlighting feature} existed as color-highlighted portions of articles; the color of the highlight denoted its most popular tag, giving an element of glanceability to the moral framing summarization of the article, as this could indicate to users if an article was relatively homogeneous in its framing (majority highlights dominated by one or a few colors), or diverse (a rainbow of highlights). 

Hovering over any highlights in an article triggered an inline pop-up to appear---the \emph{annotation feature} (Figure~\ref{panov1}, left). The annotation displayed the moral value framing \emph{tagging feature} where users could add tags to a highlight, up-voted tagged frames,
see definitions of those moral values, see \emph{comments} below each tag, and leave one's own comments (figure~\ref{panov1}, right).

In-page highlighting and annotation employed interactive methods to educate users to (i) become familiar with the moral foundations framework, and (ii) collaboratively annotate and discuss news in the context of moral frames. Highlighting and annotating text with moral frames required a deeper reading of the content, pushing the user to be conscious about moral framing. Users thus had to discover how to identify moral framing for themselves; repeated exposure to both the process of self-annotation and observation of others' annotations further reinforced the meaning of the moral frames.

\subsubsection{Extension pop-up interface}
\begin{figure}
\begin{center}
\includegraphics[height=0.55\columnwidth]{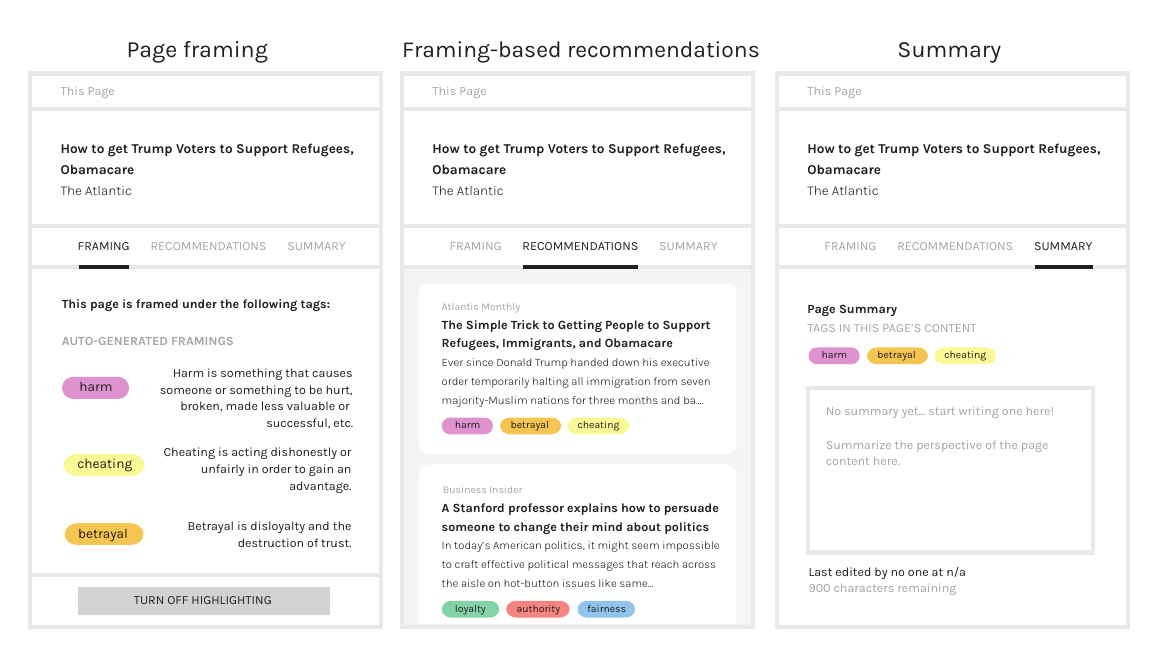}
\caption{The extension tabs. From left to right: the page framing, recommendations, and summary tabs.}
\label{extension-interface}
\end{center}
\end{figure}
The other UI \textcolor{added}{feature} was a pop-up extension interface. The interface was accessed from the Chrome browser toolbar by clicking on the extension's icon. The extension pop-up interface had three features.

\begin{figure}
\begin{center}
\frame{\includegraphics[width=0.35\columnwidth]{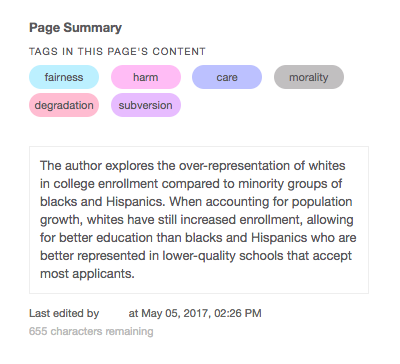}}
\caption{A summary left on a pro-affirmative action article by a user (username redacted) during the user study.}
\label{summary}
\end{center}
\end{figure}

\begin{enumerate}
    \itemsep0em 
    \item \textbf{Page framing summary}. The framing tab showed which tags (both auto-generated and user-generated) were present in the current page. Users could also toggle the in-page highlighting feature here.
    
    \item \textbf{Framing-based \textcolor{added}{article} recommendations}. The recommendations tab showed pages similar to the current page, such as articles on the same topic, with the moral value framings of those articles displayed with each recommendation.
 
    \item \textbf{Wiki-like collaborative summary}. The summary tab had a wiki-like globally visible and editable summary for the current page as well as the moral value tags for the page. Figure~\ref{summary} shows a  summary as written by a user during the field study.
\end{enumerate}

\subsection{Evaluation 1: V1}
We conducted an 8-day field study to evaluate the effectiveness of this first design. 

From participant behavior and qualitative results during the study, we gained insight into the following research questions:
\begin{itemize}
    \item \textbf{RQ1:} Did using the features affect how users approached thinking about and discussing political issues?
    \item \textbf{RQ2:} Did users engage more with certain features over others?
    \item \textbf{RQ3:} Could usage of these features lead to greater empathy for others, especially those with views opposite from one's own?
\end{itemize}

\subsubsection{Participants}
We recruited participants who read online news through university-affiliated mailing lists and social media. Potential participants filled out an interest survey and were filtered for the following criteria: (1) Google Chrome users, and (2) readers of at least 6+ online news articles weekly. 42 people completed two pre-study surveys and officially enrolled in the study. During the study period, participants read articles while using the system during an 8-day period, then filled out a post-study survey. By the end of the study period, 35 participants completed the study; the remaining 7 dropped out during the study period. They were compensated \$40 for completing the study. 

Out of the 35 participants, 51.4\% were male and 48.6\% were female, with an age range of 19-56 and an average age of 26.6. Participants were also asked to rank where they generally fall on the left-right spectrum (1 = very conservative, 2=conservative, 3=slightly conservative, 4=moderate, 5=slightly liberal, 6=liberal, 7=very liberal), with the average being between slightly liberal and liberal (5.21/7). \textcolor{added}{Given prior research demonstrating differing moral foundations between liberals and conservatives in the U.S.~\cite{mft2}, we collect political typology to roughly approximate the diversity of moral foundations in our study population. Our users skewed left, so we ensured that right-leaning users were distributed across all experimental condition.}

\subsubsection{Procedure}
There were two system conditions and two moral foundations theory (MFT) education conditions. The two system conditions were: (i) our system, the experimental condition; and (ii) the control condition, with no highlighting, annotation, or framing-related extension features, only recommended pages. The two education conditions were: (i) no MFT education, and (ii) MFT education, via an explanation of their results from taking a survey about their own moral foundations. Participants were randomly assigned to one of three groups:

\begin{itemize}
    \itemsep0em 
    \item \textbf{G1: Control.} Control system + no MFT education
    \item \textbf{G2: Education.} Control system + MFT education
    \item \textbf{G3: Treatment.} Experimental system + MFT education
\end{itemize}
We left out the condition of experimental system + no MFT education given that someone without MFT education would not be able to correctly use our tool which posited a basic understanding of MFT. Group sizes in this study were $G1 = 10, G2 = 9,$ and $G3 = 14$.

The user study was conducted virtually, with communication with participants done over email. We selected four current controversial issues in U.S. politics to be the content focus of the study. These four issues were:
\begin{enumerate}
    \itemsep0em
    \item Euthanasia/Physician-assisted suicide (PAS)
    \item Healthcare (HC)
    \item Free speech on college campuses (FS)
    \item Affirmative action in college admissions (AA)
\end{enumerate}

\textbf{\emph{Pre-study surveys}}. At the beginning of the study, all participants completed two surveys: a ``pre-study survey'', and a ``Moral Foundations questionnaire''. The pre-study survey included questions around political opinions on the pre-specified issues, how participants explained stances on political issues, and their general and issue-specific empathy levels. The issue-specific empathy questions asked participants to rank how closely statements such as ``I can see why...'' described them. The political opinions on pre-specified issues section asked participants to write ``an argument directed toward someone who disagrees with their view''. The Moral Foundations questionnaire, taken by all participants but with an extra module for MFT education in both participant groups G2 and G3, is described in the following section.

\textbf{\emph{MFT education}}. The ``Moral Foundations questionnaire'', adapted from the ``Moral Foundations Questionnaire'' by Jonathan Haidt and Jesse Graham~\cite{mft-survey}, gathered data on participants' moral foundations. We note that we found the same correlation with liberal/conservative political stance vs. moral frames in our participants that prior MFT research~\cite{graham2012moral, mft2} also demonstrated, which was that Moral Foundations of liberal-identifying persons spiked in fairness and care, while conservative-identifying persons were more even across foundations.

Upon submitting this survey, participants in the ``no MFT'' condition were shown a ``Thank you for completing this survey'' screen. Participants in the ``MFT education'' condition were shown a page detailing MFT, their results and how to interpret them, a graph of how they compare to typical \textcolor{added}{U.S.} liberals/conservatives, and what the implications of their results mean. This information also came from the Moral Foundations Questionnaire \cite{mft-survey}.

\textbf{\emph{Reading period}}. After completion of the pre-study surveys, the reading period began: participants were sent article briefings every two days, for a total of four article briefings over eight days. Each article briefing consisted of two articles about the same issue but from opposite ideological sides. \textcolor{added}{Each pair of articles per issue were chosen from popular news outlets and verified to take the appropriate common affirmative or negative viewpoint on the issue}. The articles given are shown in Table~\ref{table:articles1}. All participants were asked to read both articles and share them if they'd like. Participants in the treatment system condition (treatment group) were tasked to use all features in the system: 
\begin{itemize}
    \itemsep0em
    \item Contribute at least two observations to the page, by either highlighting and annotating sentences or contributing to an existing highlight annotation. 
    \item Read the framing tab of the extension
    \item Read and contribute to the summary tab of the extension
    \item Browse the recommendations in the recommendations tab of the extension and click on any that interest them.
\end{itemize}
Participants in the control system condition (control and education groups) were tasked to just look at the recommended pages in the extension, and click on any that interested them, as well as share either of the articles if they wished. After completion of each reading and its tasks, all participants filled out a form asking for a reflection of their opinions on the article and thoughts on the author's perspective in each article. 

\begin{table}[htbp]
    \centering
    \small
    \begin{tabular}{ |p{2cm}||p{8cm}|p{3cm}|  }
    \hline
    \multicolumn{3}{|c|}{\textbf{Articles in Evaluation 1}} \\
    \hline 
    \textbf{View} & \textbf{Title} & \textbf{Source} \\
    \hline
    Liberal & Doctor-assisted dying: The right to die & 
    The Economist \\
    \hline
    Conservative & Physician-assisted suicide is always wrong & Newsweek \\
    \hline 
    Liberal & Republicans have to pass their health care bill to find out what's in it & Reason \\
    \hline
    Conservative & My son has a preexisting condition. He's one of the reasons I voted for the AHCA. & The Washington Post \\
    \hline 
    Liberal & Campus must prioritize safety of marginalized over free speech & The Daily Californian \\
    \hline
    Conservative & When free speech becomes a free-for-all, democracy loses & The Hill  \\
    \hline
    Liberal & The best new argument for affirmative action & The Atlantic \\
    \hline
    Conservative & The painful truth about affirmative action & The Atlantic \\
    \hline
    \end{tabular}
    \caption{The articles participants read during evaluation 1. Participants were given both a liberal and conservative article on the same topic at a single time.}
    \label{table:articles1}
\end{table}

\textbf{\emph{Post-study surveys}}. At the conclusion of the reading period, participants completed the post-study survey, which was almost identical to the pre-study survey. Participants also completed open-ended questions about their experience and usage of the system.

\subsubsection{Analysis methods}
As mentioned above, in the pre- and post-study surveys, participants were asked to write ``an argument directed toward someone who disagrees with their view''. To determine whether people changed the way they framed each of their pre (baseline) and post (final) survey responses, the first two authors, blind to the group membership of subjects, independently looked for the presence of moral framing in both. Then, participants' baseline and final responses were compared against each other, and coders looked to see whether there was a change in the moral framing used. Finally, the coders' results were compared against each other, and overlapping results were kept.

There were two types of changes the coders considered positive in the moral framing of participants' responses:
\begin{enumerate}
    \itemsep0em
    \item No moral framing present to moral framing present.
    \item Arguments with moral framing in one's own frames to arguments with moral framing in an audience holding the opposing view's frames.
\end{enumerate}

\begin{table}[htbp]
    \centering
    \small
    \begin{tabular}{ |p{2cm}||p{5.4cm}|p{5.4cm}|  }
    \hline
    \multicolumn{3}{|c|}{\textbf{Moral frames in selected issues}} \\
    \hline
      & Left (liberal) & Right (conservative) \\
    \hline
    Physician-assisted suicide (PAS) & 
    Fairness: provide the choice to end one's life \newline 
    Care: opportunity to end one's suffering \newline & 
    Morality: the immorality of ending one's life\newline
    Sanctity: the sanctity of life\\
    \hline
    Health care (HC) & Fairness: opportunity for HC for all \newline
    Care: the advantaged take care of the disadvantaged 
    & Fairness: the unfairness of paying for someone else \\
    \hline
    Free speech (FS) & Care: protection from harmful speech & Fairness: the right to speak as one likes \\
    \hline
    Affirmative action (AA) & Fairness: level the playing field for the disadvantaged \newline
    Care: overcome harm from past injustices & Fairness: unfair for one to receive preferential treatment due to race alone \\
    \hline
    \end{tabular}
    \caption{Rubric of moral frames in selected issues used in moral framing analysis for evaluation 1}
    \label{table:frames}
\end{table}

For our analysis, we chose to focus on a single topic across all participants due to the time required to code statements for moral framing. We additionally chose to focus on physician-assisted suicide (PAS) as our topic of analysis because the moral framings of the left vs right arguments are distinct: liberal arguments use fairness and care; conservative arguments use morality and sanctity. PAS arguments also touch upon the most number of moral values. In contrast, as seen in the rubric, other topics such as health care and affirmative action have fairness arguments in both the left and right stance. Table \ref{table:frames} shows the general rubric used for coding moral frames. 

\subsection{Results - V1 evaluation}
\textbf{RQ1: Effect on how users approached thinking about and discussing political issues}
A majority of system users (8 of 14) indicated in the post-study survey that using the features changed their thinking on the articles and issues they read, with many users (5) mentioning moral frameworks explicitly. Multiple participants said that interacting with the features raised their awareness of how multiple framings could be used:
\begin{quote}
    \emph{``It made me realize that multiple framings can be applied to the same statements, and made me realize how many perspectives and meanings can be embedded into a certain statement. These meanings aren't always intended by the writer too---the readers and their experiences and perspectives influence the framings identified in each article.''}
\end{quote} 

\begin{quote}
    \emph{``Knowing that there were several ways a moral argument could make its point was useful.''}
\end{quote}

Participants also mentioned a newfound awareness of how authors and other readers utilize framing:
\begin{quote}
    \emph{``It made me think about the values each author had and was trying to highlight in their article.''}
\end{quote}

\begin{quote}
    \emph{``It forced me to view arguments in a new way since I don't usually consciously think about the moral frameworks they use. It was most interesting to see how other people tagged the arguments.''} 
\end{quote}

\begin{quote}
    \emph{``I think I started to examine how the article presented the argument instead of the content of the article alone.''}
\end{quote}

Additionally, 8 participants demonstrated positive improvement (\textcolor{added}{determined manually by trained coders using methodology explained} in the analysis methods section) in re-framing their responses to the following prompt on physician-assisted suicide: ``Write an argument directed toward a person from the opposite side to convince them of your side.'' Of these eight participants, five were from G3 (36\% of G3), two from G2 (22\% of G2), and one from G1 (10\% of G1). Table~\ref{reframed} shows an example of a participant's re-framed response and the moral frames present before and after.\\

\begin{table}
    \centering
    \scriptsize
    \setlength{\tabcolsep}{1em} 
    \begin{tabular}{ p{5cm} | p{5cm} }
        \textbf{Pre-study} & \textbf{Post-study} \\
        & \\
        ``If the goal is compassionate care and easing the suffering of terminally ill people, then we can agree that if someone is suffering than they should have the right to end their pain.`` & ``Human life is extremely valuable but so is the person's autonomy and dignity. If someone decides they are suffering so much they do not want to live anymore, we can actually show more care towards this person by honoring their wishes instead of letting them continue to suffer.``\\
        & \\
        Care & Sanctity, care \\
    \end{tabular}
    \caption{Pre- and post-study survey response from a participant demonstrating re-framing into the moral values of the audience (a conservative).}
    \label{reframed}
\end{table}

\begin{table*}
    \centering
    \scriptsize
    \begin{tabular}{ l r r r r r r  }
    \toprule
       Briefing & \makecell{Highlights/article} & \makecell{Avg tags/ highlight} & \makecell{Avg votes/tag} & \makecell{Summary edits} & \makecell{Comments/tag} & 
       \makecell{Avg interactions/user} \\
    \midrule
    1. PAS - left & 26 & 1.7692 & 2.1087 & 1 & 0 & 6.3333 \\

    1. PAS - right & 36 & 1.5278 & 2.1818 & 1 & 0.0727 & 7.6667\\
    \hline
    2. FS - left & 19 & 1.6842 & 2.1563 & 1 & 0.0313 & 3.1429 \\

    2. FS - right & 17 & 1.2941 & 2.0909 & 1 & 0.0455 & 5.1538 \\
    \hline 
    3. HC - left & 20 & 1.45 & 2.2414 & 1 & 0.1034 & 4.7690 \\

    3. HC - right & 21 & 1.3810 & 2.3103 & 1 & 0 & 4.9231 \\

    \hline
    4. AA - left & 15 & 1.5333 & 2.2609 & 1 & 0.0435 & 3.7143\\

    4. AA - right & 22 & 1.4545 & 2.0312 & 1 & 0 & 4.6429 \\
    \bottomrule
    \end{tabular}
    \caption{Table of user engagement with system V1 features}
    \label{table:desc}
\end{table*}

\textbf{RQ2: Did users engage more with certain features over others?}
A number of measurements around user engagement with the system were taken throughout the user study. Table \ref{table:desc} shows these various metrics. These metrics show high levels of engagement for some components, such as highlights per article, and low levels of engagement for others, such as summary edits. Highlighting was a feature that was heavily engaged with---users left many highlights, with each highlight followed up with at least one tag. Comments per tag, however, were not utilized, indicating to us that users did not find commenting at the tag level useful. Summary edits was consistently at just 1 for each article---we observed that once the first person made the initial summary, no other users followed up with edits. Empirically, users noted in post-study feedback that being able to see others' highlights and tags was valuable: \emph{"It was cool for me to see the sorts of themes that other people picked out"; "it was most interesting to see how other people tagged the arguments."}.

Of the three features in the extension pop-up, users were most interested in the Recommendations feature and not interested in the other two, with multiple (4) participants mentioning they explicitly liked the Recommendations tab when prompted about their thoughts on the extension features: \emph{"I really liked the recommendations tab"; "My favorite would be "Recommended Pages"; "I thought the recommendations tab was pretty cool"}. Users did not particularly like or find value out of the other two features, with multiple participants commenting that they did not use them.

We observed through both qualitative feedback and user behavior that users wanted to discuss and debate. Although users did not comment at the granular level of the tag, we observed some users taking advantage of the bulletin board feature on the extension's default view which allowed for leaving general comments on a page as a whole. We did not prompt users to interact with this feature in the study. Users naturally left their opinions on the issues in articles without any kind of prompting from us or the extension. Users also mentioned in the post-study survey feedback section that they'd be interested in more developed commenting functionality, different from the way it was implemented in V1: 

\begin{quote}
    \emph{``Would have been more interesting if people's comments were front and center.''}
\end{quote}
\begin{quote}
    \emph{``I do like the idea of letting people comment on the tags but I'm not sure how conducive it is to have a full discussion just in that thread.''}
\end{quote}

We use these results and metrics to inform the design of the second design iteration.\\
\\

\textbf{RQ3: Could usage of the features lead to greater empathy for others, especially those with views opposite from one's own?}
Finally, we looked at participants' self-reported empathy on the perspectives of the other side for the selected issues. This was measured using the empathy survey included in the pre- and post-study surveys as previously described. Overall, the positive change in empathy was greater for system users compared to the control, but the results were not statistically significant---a 23\% vs 15\% increase in average score on the empathy survey. 





\section{Iteration 2: V2}
In this section, we describe the design of the second design iteration of features and a field study evaluation.

\subsection{System Design}
\begin{figure}
\begin{center}
\frame{\includegraphics[width=0.47\columnwidth]{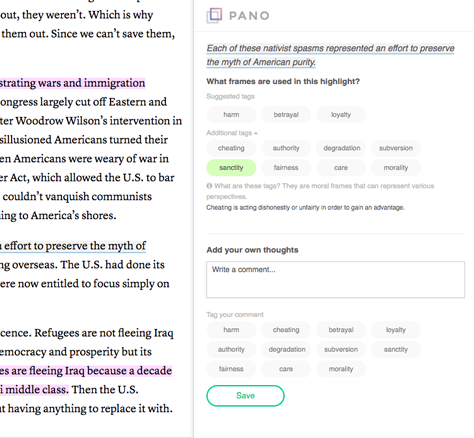}}
\hspace{0.2cm}
\frame{\includegraphics[width=0.47\columnwidth]{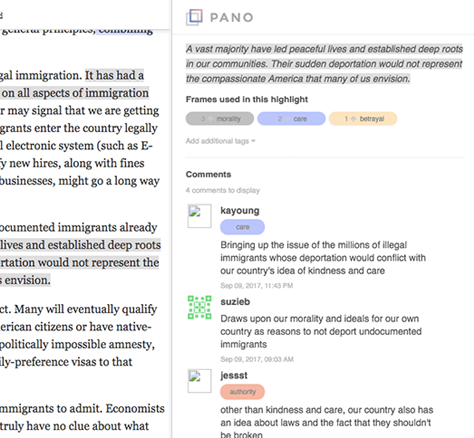}}
\caption{V2, the second iteration. Left: add annotation interface; right: annotation interface.}
\label{panov2}
\end{center}
\end{figure}

The second design iteration of features for engaging with news using moral framing was motivated by our learnings from the first field study. As we discovered, users were eager to discuss issues and leave their own opinions on articles; they also much preferred the in-page highlighting and annotation over the extension pop-up features that were separate from the article content. Therefore, the design of the second version focuses on the collaborative highlighting and annotation component, though we keep the extension pop-up features from before since some users found the recommendations useful.

Compared to the first version, the second version \textcolor{added}{focuses on commenting and discussion as the core feature and interaction mode}. As before, users highlight text in articles to trigger an interface where they can tag the moral frames present in the text. This time, users can also leave a comment simultaneously, and tag the comment with what moral frames they are speaking from. Drawing on Premack's principle~\cite{premack}, or using more preferred behaviors as reinforcement for less preferred behaviors, subsequent users who wish to respond to a comment on that annotation or leave their own thoughts must demonstrate that they also understand and have examined the moral framing in that annotation. They must contribute to the tagging on that annotation by either upvoting existing tags, or adding tags they believe are missing. Only after providing such a contribution is the commenting feature unlocked. Figure~\ref{panov2} shows the V2 user interface.

\subsection{Evaluation 2: V2}
Similar to the first iteration, a 6-day field study was conducted to evaluate the effectiveness of this second design. We looked for the same things as before: effect on how users approach issues (RQ1) and change in empathy for others on issues (RQ3). We did not compare relative engagement between all the different system features in V2 since this iteration homed in on the commenting and annotation features.

\subsubsection{Participants}
As in Study 1, we recruited participants who read online news through university-affiliated mailing lists and social media and filtered for the same criteria. 36 people completed two pre-study surveys to officially enroll into the study. During the study, participants read articles while using the experimental system over a 6-day period, then filled out a post-study survey. By the end of the study period, 24 participants completed the study and were compensated \$30 for completing the study. We note that the dropout rate was largely due to the study tasks being involved and time-intensive, and that dropout rates between all study condition groups were approximately equal.

\subsubsection{Procedure}
The procedure for Study 2 was similar to Study 1, but with a few major changes. There were two groups this time, with participants randomly assigned to one of the groups. Group sizes in this study were $G1 = 13$ and $G2 = 11$.
\begin{itemize}
    \itemsep0em 
    \item \textbf{S2.G1: Control.} Control system + no MFT education
    \item \textbf{S2.G2: Treatment.} Experimental system + MFT education
\end{itemize}

Instead of four current issues as in evaluation 1 for the content focus of the study, we narrowed in on one---immigration, for which we identified three different points: (i) the general immigration debate; (ii) Deferred Action for Childhood Arrivals (DACA); (iii) refugees.

At the beginning of the study, all participants completed a pre-study survey and the Moral Foundations Questionnaire as described in Study 1's procedure. Then, the reading period began: participants were sent article briefings every two days, for a total of three article briefings over six days. The articles given are shown in Table~\ref{table:articles2}. Each article briefing was structured the same as in Study 1. Participants in the treatment group were tasked to use the system features:

\begin{itemize}
    \itemsep0em 
    \item Contribute at least two observations to the page. This can be done by either of the following.
    \begin{itemize}
        \itemsep0em 
        \item Highlight/tag a sentence and leave a comment. 
        \item Comment on something someone else highlighted. In order to be eligible to do so, users first need to contribute to the tagging of that highlight, by upvoting any existing tags or adding ones they think may be missing.
    \end{itemize}
\end{itemize}

\begin{table}[htbp]
    \centering
    \small
    \begin{tabular}{ |p{2cm}||p{8cm}|p{3cm}|  }
    \hline
    \multicolumn{3}{|c|}{\textbf{Articles in Evaluation 2}} \\
    \hline 
    \textbf{View} & \textbf{Title} & \textbf{Source} \\
    \hline
    Liberal & Why immigration is good: 7 common arguments against reform, debunked & Bustle \\
    \hline
    Conservative & The immigration debate we need & 
    The New York Times \\
    \hline 
    Liberal & Ending DACA would be mean-spirited and shortsighted – even for Trump & Los Angeles Times \\
    \hline
    Conservative & Rescinding DACA is the right thing to do & The Federalist \\
    \hline 
    Liberal & What America Owes Refugees from the Middle East & The Atlantic \\
    \hline
    Conservative & Refugee Madness: Trump Is Wrong, But His Liberal Critics Are Crazy & The National Review  \\
    \hline
    \end{tabular}
    \caption{The articles participants read during evaluation 2. Participants were given both a liberal and conservative article on the same topic at a single time.}
    \label{table:articles2}
\end{table}

Participants in the control group were tasked to just read the articles. After completion of reading and tasks, all participants filled out a reflection form. At the end of the reading period, participants filled out the post-study survey and answered open-ended questions about their experience.

The data collected from this field study was analyzed using the methods outlined for the first field study of V1. Table~\ref{table:frames2} shows the rubric used for coding moral frames. All user responses to the pre- and post-study re-framings of stances on political issue questions were coded.

\begin{table}
    \centering
    \scriptsize
    \begin{tabular}{ |p{2cm}||p{5.4cm}|p{5.4cm}|  }
    \hline
    \multicolumn{3}{|c|}{\textbf{Moral frames in selected issues}} \\
    \hline
      & Left (liberal) & Right (conservative) \\
    \hline
    General immigration & 
    Fairness: America was founded by immigrants, people should be given a chance \newline
    Care: immigrants come to America to seek a better life & 
    Cheating: illegal immigration is unfair to legal immigrants \newline
    Authority/subversion: illegal immigrants subvert the law \newline
    Degradation: immigrants degrade American culture by not assimilating \\
    \hline
    Deferred Action for Childhood Arrivals (DACA) & 
    Fairness: People affected did not choose to come here\newline
    Care/harm: deportation to a country not their own is cruel\newline
    Betrayal: repealing DACA reneges on an Obama-era promise &
    Subversion: the original DACA was unconstitutional\\
    \hline
    Refugees & 
    Care: refugees are suffering in their home countries \newline
    Morality: we have a moral responsibility to help the less fortunate
    & Loyalty: protect our own citizens first \newline
    Degradation: refugees will bring crime and terror \\
    \hline
    \end{tabular}
    \caption{Rubric of moral frames in selected immigration issues for evaluation 2}
    \label{table:frames2}
\end{table}

\subsection{Results}
\label{results}

\textbf{RQ1: Effect on how users approached thinking about and discussing political issues}
The vast majority of experimental system users (9 out of 11) indicated that using the features affected the way they thought about articles. These users indicated that it helped reveal moral framings and biases while reading the articles, and in some cases, even helped them discover why \textcolor{added}{people on opposite stances of an issue} talk past each other.
\begin{quote}
    \emph{``It did force me to think about the different moral perspectives/priorities of the views, which revealed parts of why there was disagreement between the two sides.''}
\end{quote}
\begin{quote}
    \emph{``Sometimes [the system] made me realize that the same statement was interpreted very differently by another person.''}
\end{quote}

Many users $(n=7)$ mentioned how using the features caused them to give more thought to the way they perceived the authors or other commenters, helping users put a more fundamentally human element into perspectives and authors. 
\begin{quote}
    \emph{``It forced me to look beyond my first impression on the article and analyze the piece more objectively. It makes you acknowledge the people and the ideas for or against the article you're reading. It definitely encourages political conversation, and makes people start talking about what really matters.''}
\end{quote}
\begin{quote}
    \emph{``It helped put a lot of things in perspective, and helped show how there's always more to the story than you see in a single statement by one person.''}
\end{quote}

\textbf{Experimental system users utilized moral framing of the other side in comments.}
On each article in the study, we observed cases of users writing comments explicitly tagging and framing with the moral values of the side opposite their perspective. For instance, on a pro-Deferred Action for Childhood Arrivals (DACA) article, users framed arguments directed toward anti-DACA readers in terms of conservative values on immigration such as loyalty (to one's country) and authority (the power of the law). Figure~\ref{results} shows one such example.

\begin{figure}
\begin{center}
\frame{\includegraphics[width=0.4\columnwidth]{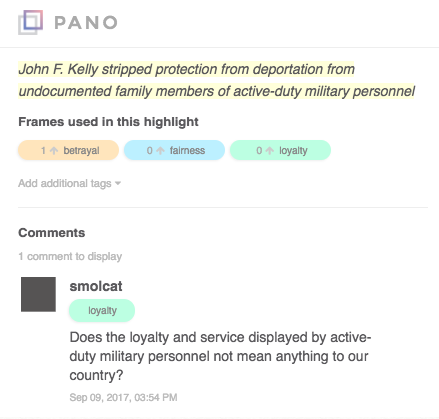}}
\caption{A comment from a user during the field study demonstrating framing a pro-immigration argument directed at an anti-immigrationist using anti-immigration moral framing.}
\label{results}
\end{center}
\end{figure}

Consider the case study of a liberal-leaning system user in our study, User X. User X demonstrated strong re-framing into the moral frames of the audience in their responses to all three of the pre- vs post-study survey questions.
For article briefing \#1, user X started off using only liberal moral framings (fairness, care) in their comments; however, in the latter two article briefings, user X started utilizing a more diverse set of moral framings, including conservative moral frames (loyalty, authority).

These provide positive examples of how we imagined our tool being used. Our results demonstrate the promise of tools such as ours for encouraging reflection about the underlying moral values behind different ideological stances, towards the ability to both recognize and employ framing in argumentation, as well as gain empathy for diverse perspectives.\\
\\
\textbf{RQ3: Could usage of the features lead to greater empathy for others, especially those with views opposite from one's own?}
Experimental system users demonstrated greater change in self-reported empathy on the immigration issues covered in the field study when comparing pre- and post-study survey results against the control group that did not use the system ($\mu = 0.25, p < 0.05$, permutation test with differences in final vs. baseline empathy). The empathy score scale ran from 1--5. System users also had greater change in self-reported general empathy; the results are statistically significant at the 90\% confidence level ($\mu = 0.21, p = 0.08$, permutation test with differences in final vs. baseline empathy).



\section{Discussion}
Our work designing features for bridging the ideological divide explores a new approach centered around getting users to engage with framing and Moral Foundations Theory to better understand the root of issues and how others approach them. An informed society with both exposure to diverse perspectives \textit{and} the tools to productively engage with diverse perspectives is important, and our work explores designing features to position people to be better equipped to receive and engage with diverse perspectives when they encounter them. The suite of features we developed aims to help users think about issues more fundamentally, using the language of shared human values to build empathy, any time they read news by introducing moral framing-centric interactions into common actions users are familiar with. Highlighting, tagging, commenting, and recommendation systems are features present in many of the everyday digital tools and apps we use. This work showed how integrating the element of moral framing into such pre-existing systems when engaging with news can be a promising direction.

In our evaluations, we found that using certain features such as moral framing-centric highlighting and commenting helped encourage people to think in a moral framework while reading news---users demonstrated ability at identifying moral framing in specific points made by articles, as well as in using moral framing in their own comments responding to articles. Illumination of frames allows people to easily visualize similarities and differences and use this information while conversing and reading. In our commenting and highlighting features, users observe other users' moral value tags and encounter one of two scenarios: (i) they encounter moral values matching their own internal values, or (ii) they encounter moral values incongruous with their own. In our studies, both of these scenarios were valuable---the former established common ground with others, who may or may not align with the same ideological lines, and the latter helped users realize \emph{why} the ideological divide exists, at least on specific issues.  As one user from Evaluation 2 said:
\begin{quote}
    \emph{``Sometimes it was very frustrating to believe that people thought that way. Sometimes it was very lovely to know someone else thinks the way you do. And other times it was enlightening to look at something perhaps from a way you'd never expect.''}
\end{quote}

Finding ways to surface frames in various features while interacting with news also provides an interesting way to approach teaching media literacy. In conjunction with past work on media frames as discussed in \ref{prior-work-framing}, bringing principles like visualization of media frames and language used~\cite{diakopoulos2015compare, mediacloud} to everyday news audiences provides a way for users to gain a deeper awareness to their engagement with content. 

Our studies explored the benefits of both annotation and presentation of annotation. Its methods of user engagement with moral framing in news drew from the principles of Bloom's taxonomy, utilizing the importance of having users carry out the three most effective methods of learning: \emph{analysis}, \emph{evaluation}, and \emph{creation}~\cite{bloom}. In using each feature, users needed to analyze presented perspectives for the appropriate moral frame tags, evaluate the correctness of existing tags, and create by composing comments focused around moral framing. In our evaluations, users were creating arguments on physician-assisted suicide, health care, affirmative action, free speech (Study 1), and immigration (Study 2) for an audience from the other side. The comment-in-the-margin design limited the scope of conversation to the corresponding highlighted point. By constant association and reminder of framing through visual cues such as the tags and highlight color, the design reinforced the link in users' minds between the perspective being presented and the underlying values it represents. 

Explorations like ours are a step toward higher-quality engagement with news and more productive online discussions. We designed our features such that users can only leave comments and participate in existing discussions on an annotation if they have contributed to that annotation via tagging moral values or supporting (up-voting) existing moral values. Users, therefore, are demonstrating a level of understanding, or at the very least, consideration for, the moral values present in the issue being discussed. In addition, prompting users to re-frame issues in different moral frames could promote active listening like with the Reflect tool~\cite{Kriplean:2012:TYM:2207676.2208621}. 

\textcolor{added}{Our studies suggest that certain features that help audiences engage with moral framing in news and online discourse show promise towards bridging ideological divides.} We consider how these designs can be integrated into the ways in which people today already engage with news through existing systems. Many of these features can exist on surfaces that are already a part of users' everyday routines for news consumption through social media, online forums, and web browsing. From our learnings with recommendation features, examples of ways to integrate moral framing include generating article recommendations that factor into account a user's values for diverse but value-aligned content. From our learnings with highlighting, commenting, and tagging features, examples include highlighting moral framing while engaging with media in news feeds, centering framing or tagging around themes like moral foundations in comments sections in online discussions, providing re-framed, and personalized news such as properly-framed headlines. 

\textcolor{added}{However, it's important to note the context of such social annotation and engagement systems. As is commonly known and well-studied in past work, online social spaces open to the general public are oftentimes unproductive, especially open comments sections, and prior work has suggested incivility online contributes to further polarization~\cite{nasty}. In light of this knowledge, the direction of our work might be more successfully deployed in controlled environments, such as in smaller, well-moderated community groups or educational forums. Such forums could be led by educators in a formal classroom or more informally, such as in the case of publisher-led communities on Facebook Groups~\cite{vox}. This is especially motivated by our promising results with users recruited from higher-education channels interested in news who may be more politically-aware than the average person, as these demographics may be more likely to be present in those settings. Given prior work discussed in \ref{social-annotation} on social annotation as an effective tool for engaging with content in classrooms, we also believe educational contexts are an promising area to explore introducing Moral Foundations-based interventions.} Future work could explore integrating these designs into such existing systems.





\section{Limitations and Future Work}
We now discuss some of the limitations we encountered in the process of developing and evaluating our designs, as well as promising directions for future work. First, we acknowledge limitations in our study sample population.  \textcolor{added}{Our user recruitment took place through university-affiliated channels, which meant our users are inherently skewed toward being well-educated. Additionally, our group sizes were limited to a relatively small set of participants, insufficient to achieve statistical significance with any quantitative analyses.} The nature of a longitudinal field study meant that it was challenging for people to commit for the term of the study, as tasks had to be performed on a regular cadence of every two days. Second, we discovered during our user study recruitment that getting a good proportion of people who politically lean conservative \textcolor{added}{or hold traditionally conservative stances on our key issues} is hard for the authors---likely due to our own echo chambers, \textcolor{added}{and recruitment taking place within a higher education context}. The majority of respondents to our recruitment surveys were moderate-to-liberal. We attempted to mitigate the imbalance of conservatives in our second user study by publicizing out to a wider net and more diverse universities; however, the number of conservative participants still remained low. 

\textcolor{added}{Another limitation of this work is that in our user study for V2, we did not collect data from control group \textbf{S2.G1} on whether simply participating in the study (i.e., reading articles and reflecting on them) changed the way they thought about the articles. We collected participant responses to this question only for the treatment group, \textbf{S2.G2}, and those responses informed the results of \textbf{RQ1} in \ref{results}.}

In analyzing results from the user study, we needed to extract moral framing from hundreds of free-response answers. Our first idea was to automate the process using the LIWC methodology explained in \ref{inpagehighlight} that was used within the annotation system for generation of recommended tags; however, we quickly discovered through early trials that LIWC was not accurate enough for this use case to aid analysis---responses run through LIWC yielded noisy results because the individual inputs were small (on average a few sentences long). Given its intended purpose for longer text passages, rather than the shorter format of our data, we decided to go with the manual approach. Thus, we coded hundreds of free-response answers manually for usage of moral framing. This was both time-intensive and difficult, as framing could often be ambiguous. We mitigated the variance in coder analyses by having multiple coders for all responses and took the intersection of framings identified.

\textcolor{added}{Given the limitations outlined above, we believe that a promising direction for future work in addition to design directions discussed in the previous section could be to deploy these features to controlled environments, for instance, a classroom environment for educational purposes. A quantitative study on the effects of these MFT-based interventions could be done with a larger group with a more ``in-the-wild'' approach, by providing the system to such an audience and passively observing user interactions over a period of time. In such a study, a control group using a similar system but with only the social features and not the MFT-based interventions on top could be established to use as a baseline for comparison. In addition, a different recruitment strategy could allow for studies of the effectiveness of our designs with participants who are less well-educated or politically-aware.}

Another direction for future work is to address the cold-start problem  users of certain features may encounter. It is easy to imagine the experience of using highlighting and commenting features on a popular article along with a number of other users; however, edge cases of the experience include being the first user on an article, or using these features on an unpopular article. In these cases, it is important to ensure that the features can still provide the necessary utility toward achieving their goals and giving these types of users opportunities to interact with the moral framing in the text. Improvements to make the automatic moral frame detection and extraction algorithm more accurate can help mitigate cold-start problems by properly pre-populating articles with highlights and tags as if it had been done by other users.

\section{Conclusion}
In today's political climate, ideological divides are widening. To address this, we contribute the design and evaluation of a suite of features for online systems to engage with news using the Moral Foundations Theory framework. In this paper, we presented two iterations of design and evaluation of various features, \textcolor{added}{from which we found evidence that suggests user improvement in empathy and ability to re-frame arguments effectively}, as well as more generally, changes in the way they thought about news and selected issues with regards to awareness of moral framing. We believe that our findings and designs could be useful in further work on using framing to encourage productive interactions with news and online political discussions, especially in providing insight into how to introduce engaging with moral framing into everyday systems.

\received{July 2021}
\received[revised]{September 2021}
\received[accepted]{October 2021}

\bibliographystyle{ACM-Reference-Format}
\bibliography{bibliography}

\end{document}